\documentclass{vgtc}                          

\ifpdf
  \pdfoutput=1\relax                   
  \usepackage{graphicx}                
  \DeclareGraphicsExtensions{.pdf,.png,.jpg,.jpeg} 
\else
  \usepackage{graphicx}                
  \DeclareGraphicsExtensions{.eps}     
\fi%

\graphicspath{{figures/}{pictures/}{images/}{./}} 
\usepackage{orcidlink}
\usepackage{microtype}                 
\PassOptionsToPackage{warn}{textcomp}  
\usepackage{xcolor}
\usepackage{textcomp}                  
\usepackage{mathptmx}                  
\usepackage{times}                     
\usepackage{cite}                      
\usepackage{tabu}                      
\usepackage{booktabs}                  
\usepackage{enumitem}
\usepackage{changepage}

\onlineid{0}

\vgtccategory{Research}

\vgtcinsertpkg

\title{SportsBuddy: Designing and Evaluating an AI-Powered Sports Video Storytelling Tool Through Real-World Deployment}

\author{
Tica Lin \orcidlink{0000-0002-2860-0871} \thanks{
\textit{
Tica Lin and Ruxun Xiang contributed equally to this work.}
\\ 
\hangindent=4mm \textbullet\ \textit{T. Lin (mlin@g.harvard.edu), R. Xiang (ruxunx@g.harvard.edu), M. Chiang, C. Ye were affiliated with Harvard University during this work. G. Liu and H. Pfister are with Harvard University. }
\\
\hangindent=4mm \textbullet\ \textit{D. Tiwari and Z. Chen (ztchen@umn.edu) are with University of Minnesota. }
\vspace{-1cm}
} ,
Ruxun Xiang \orcidlink{0009-0000-3257-4859} \footnotemark[1] , 
Gardenia Liu \orcidlink{0009-0002-5442-9788}, 
Divyanshu Tiwari \orcidlink{0000-0003-4279-8332}, 
Meng-Chia Chiang, 
Chenjiayi Ye \orcidlink{0009-0006-0030-9296}, \\
Hanspeter Pfister \orcidlink{0000-0002-3620-2582}, Chen Zhu-Tian \orcidlink{0000-0002-2313-0612} 
}

\teaser{
\vspace{-2mm}
\begin{center}
  \includegraphics[width=\linewidth, alt={}]{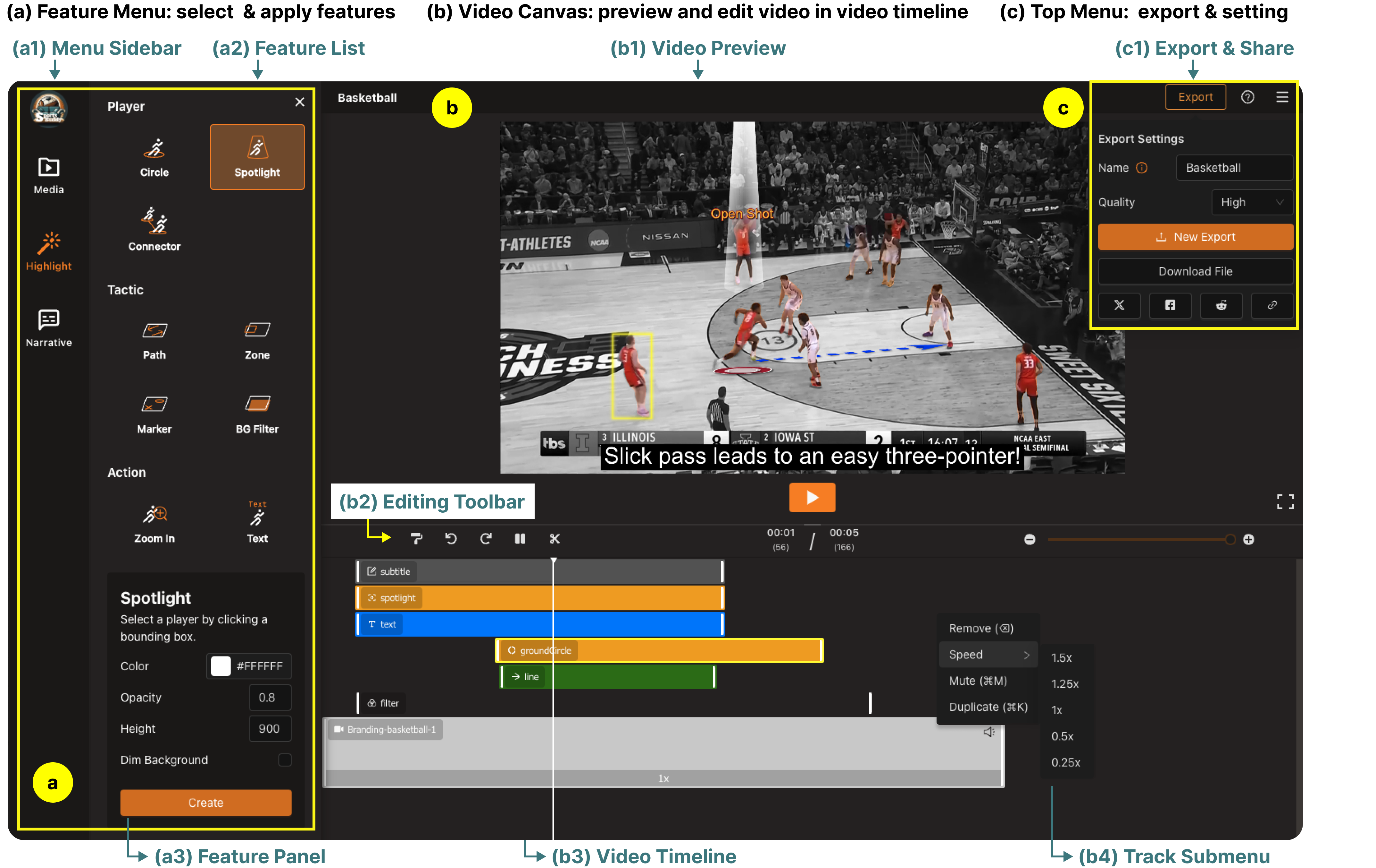}
  \caption{%
  	\SB{} enables easy highlight creation for video storytelling with intuitive visualization features and workflow. Users can (a) manage clips in \Media{} and choose from \Highlight{} and \Narrative{} features; (b) interact directly with the video canvas to apply features, adjust effects, and edit using the video timeline; and (c) share the final highlight via link or file. 
  }
  \label{fig:teaser}
  \end{center}
}

\abstract{
Video storytelling is essential for sports performance analysis and fan engagement, enabling sports professionals and fans to effectively communicate and interpret the spatial and temporal dynamics of gameplay. Traditional methods rely on manual annotation and verbal explanations, placing significant demands on creators for video editing skills and on viewers for cognitive focus. However, these approaches are time-consuming and often struggle to accommodate individual needs. SportsBuddy addresses this gap with an intuitive, interactive video authoring tool. It combines player tracking, embedded interaction design, and timeline visualizations to seamlessly integrate narratives and visual cues within game contexts. This empowers users to effortlessly create context-driven video stories. Since its launch, over 150 sports users, including coaches, athletes, content creators, parents and fans, have utilized SportsBuddy to produce compelling game highlights for diverse use cases. User feedback highlights its accessibility and ease of use, making video storytelling and insight communication more attainable for diverse audiences. Case studies with collegiate teams and sports creators further demonstrate SportsBuddy’s impact on enhancing coaching communication, game analysis, and fan engagement.
} 

\begin{document}

\newcommand{\SB}{SportsBuddy}
\newcommand{\Media}{\texttt{Media}}
\newcommand{\Highlight}{\texttt{Highlight}}
\newcommand{\Narrative}{\texttt{Narrative}}
\newcommand{\Circle}{\textit{Circle}}
\newcommand{\Spotlight}{\textit{Spotlight}}
\newcommand{\Connector}{\textit{Connector}}
\newcommand{\Path}{\textit{Path}}
\newcommand{\Zone}{\textit{Zone}}
\newcommand{\Marker}{\textit{Marker}}
\newcommand{\BGFilter}{\textit{BG Filter}}
\newcommand{\Zoom}{\textit{Zoom In}}
\newcommand{\Text}{\textit{Text}}
\newcommand{\Caption}{\textit{Caption}}
\newcommand{\revision}[1]{\textcolor{black}{#1}}

\firstsection{Introduction}

\maketitle

\firstsection{Introduction} 

Videos are essential to deliver values in the sports world. Whether through TV broadcasting, social media, film reviews, or highlight reels, they empower sports professionals to record, analyze, and refine their performance while captivating audiences for both educational and entertainment purposes.
Particularly, sports highlights—short videos capturing key moments—provide an effective format for sharing insights with a wide audience. Effective sports highlights are often enhanced with visual aids or narratives to better convey stories and guide viewer attention,
leading to growing demands on social media platforms like Youtube and Tiktok~\cite{genz}. 
However, creating sports highlights still requires extensive video editing skills, making it challenging to effectively express personal insights for the majority of sports roles, including billions of coaches and athletes globally. 

Existing solutions, such as popular video editing software like Adobe Premier Pro~\cite{premiere}, have a steep learning curve and are tedious to operate. Due to the dynamic nature of sports, narratives and focal points can shift moment by moment, requiring precise synchronization of insights and videos. As a result, content creators report spending hours to produce even minute-long highlights. 
Furthermore, making effective sports highlights often requires sports-specific graphics and animations to explain tactics and emphasize key targets, typically unavailable to non-media professionals. As a result, sports experts with deep insights but limited video creation skills still rely on manual annotations (e.g., tactic boards) and verbal explanations to convey their ideas, which limits their ability to effectively engage audiences in team or public settings.

%
To bridge the gap between sports insights and video storytelling,
prior research leverage visualization and AI techniques and propose new ways of sports highlight authoring. Most prominently, data-driven approach~\cite{chen2021augmenting} detects objects and actions in sports videos and directly map visualizations to the objects, eliminating the need for manual frame-by-frame graphic placement with traditional video editing approach. 
Intuitive interaction such as gesture and natural language could drastically ease the video authoring workflow~\cite{chen2022sporthesia,chen2023iball}. 
In addition, design framework for sports-specific visualizations were also thoroughly explored to support systematic augmented sports video creation~\cite{lin2022quest, yao2023designing}. 
Despite these recent advancements in data-driven sports video creation, there is still a lack of accessible tools for sports practitioners. Our work aims to put these visualization theories in practice to make real-world impact.

Building upon the data-driven approach in~\cite{chen2021augmenting} and embedded visualization framework in~\cite{lin2022quest}, we develop \SB{}, an AI-powered sports highlight creation tool, with intuitive web interface that allows easy integration of graphic and narratives into videos. It supports various sports, including basketball, soccer, volleyball, lacrosse, and tennis.
Our system leverages multimodal AI models to detect players and events, and provide easy addition of sports visualizations through embedded interaction (i.e. direct interaction with the objects in the video) and interactive video timeline. 
\SB{} supports an end-to-end highlight creation journey, 
including importing video to the \Media{}, adding \Highlight{} features to the video, and adding \Narrative{} with manual subtitles or AI-generated captions.
As sports practitioners create sports highlights to 1) emphasize key players,
2) illustrate tactic execution,
and 3) associate data / insights to performance, we design different visualization categories to collectively support these goals, including visualizations focusing on player, tactic, and action, as shown in Fig.~\ref{fig:teaser} (a).

Since launching \SB{} online in August 2024\footnote[1]{\SB{} can be accessed on https://sportsbuddy.online}, 
we have attracted users from a variety of sports roles, each with distinct needs for creating engaging highlights. Over 150 users have signed up, with approximately one-third being coaches focused on game analysis to improve player and team performance, another third consisting of content creators crafting video stories to engage their audiences, and the remainder comprising athletes and parents creating highlights for building profiles.
A case study with Harvard Basketball and Soccer Teams illustrated the ease of use and value of \SB{}'s sports visualization and video features. These tools empowered team marketing managers to efficiently showcase team achievements and in-depth tactical breakdowns through engaging highlights on social media. Another case study with professional and aspiring content creators emphasized how \SB{} lowers the barrier of entry for creating high-quality sports highlights, enabling new creators with limited video editing skills to master sports video storytelling.  
Continuous user feedback has also led to feature additions and usability improvements, suggesting the positive and growing impact \SB{} has in sports community.

Our paper makes four main contributions: 1) the design and implementation of \SB{} for easy sports highlight creation, 2) the integration of state-of-the-art multimodal AI to enable innovative sports video storytelling features, 3) system deployment and testing in real-world settings with active users and 4) insights into the impact of sports video storytelling derived from first-hand user evaluations.


\section{Related Work}
We review previous work on data videos for storytelling, video-based visualizations in sports, and authoring tools for data videos.

\subsection{Data Videos for Storytelling}

Storytelling plays a vital role in communicating data, providing a structured and engaging means of sharing insights~\cite{DBLP:journals/tvcg/SegelH10}. 
Data videos combine visual elements, auditory, and raw footages to create compelling stories backed by data. 
Researchers have extensively explored the design space of data videos, 
offering valuable insights for creating and authoring them.

For example, Amini et al.~\cite{DBLP:conf/chi/AminiRLHI15} analyzed 50 data videos from diverse sources and used Cohn's visual narrative structure theory~\cite{DBLP:journals/cogsci/Cohn13} to categorize these videos into four narrative structures.
Their findings unveiled design patterns and offered implications for the development of data video authoring tools.
Similarly, Thompson et al.~\cite{DBLP:journals/cgf/ThompsonLLS20} developed a design space for animated data graphics to support the development of future authoring tools. 
Cao et al.~\cite{DBLP:journals/vi/CaoDCWSZT20} analyzed 70 data videos and proposed a taxonomy for  narrative constructs, broadening our understanding of data storytelling.
More recently, researchers have explored incorporating cinematic techniques into data video design.
Xu et al. contributed a series of works, including guidelines for creating cinematic openings~\cite{DBLP:conf/chi/XuYY0WQ22}, endings~\cite{DBLP:conf/chi/XuWYWHYQ23}, and overarching structures~\cite{wei2024telling} in data videos. 
Their studies, inspired by cinematic storytelling, analyze film and data video examples to provide systematic guidance for practitioners.

While these works are relevant, sports video storytelling introduces unique challenges compared to traditional data video narratives.
Sports video storytelling focus on enhancing real-world footage with visual elements, such as overlays, annotations, and real-time data visualizations.
This setting presents a distinct design space with unique requirements and techniques.
Zhu-Tian et al.~\cite{chen2021augmenting} explored this design space, identifying six narrative orders and common usage scenarios to provide practical guidance for designing authoring tools tailored to this domain.
Building on prior work, \SB{} was developed as an augmented sports video authoring tool, inspired by insights from both traditional data video storytelling research and the unique requirements of sports storytelling.

\subsection{Video-based Visualizations for Sports}
Given the inherently visual and media-rich nature of sports, 
video-based visualizations have become a crucial format for sports data, 
as highlighted by Perin et al.~\cite{DBLP:journals/cgf/PerinVSSWC18}. 
\SB{} specifically focuses on embedded visualizations in videos~\cite{DBLP:journals/tvcg/WillettJD17}, where data visualizations are seamlessly integrated into the video as part of the scene.

In academia, a representative example is the system developed by Stein et al.~\cite{DBLP:journals/tvcg/SteinJLBZGSAGK18}. 
Their system processes raw footage of soccer games to automatically visualize tactical information as graphical marks within the video. 
Following their work,
Zhu-Tian et al. developed a series of human-AI collaborative tools that support data visualization in sports videos, ranging from racket-based sports\cite{chen2021augmenting, chen2022sporthesia} to team-based sports~\cite{chen2023iball}. 
Beyond system development, Zhu-Tian et al.~\cite{chen2021augmenting} and Lin et al.~\cite{lin2022quest} have also explored the design space of embedded visualizations in sports videos.
In industry, the strong market demand has led to the emergence of highly successful commercial systems, as pointed out by Fischer et al.~\cite{DBLP:journals/corr/abs-2105-04875}. 
For instance, Piero~\cite{piero} and Viz Libero~\cite{viz-libero} are widely used in sportscasting, offering powerful functionalities for editing and annotating sports videos. 
Additionally, Second Spectrum~\cite{secondspectrum} automatically tracks players' positions and embeds real-time status information to enhance audience engagement.
Yet, these commercial systems often cater to proficient video editors, resulting in a steep learning curve for sports analysts. 

\SB{} seeks to bridge the gap between academic research and industry by developing an easy-to-use tool targeted at real-world users. 
By simplifying workflows and making advanced video-based sports visualizations accessible, 
\SB{} aims to transition academic insights into practical solutions for broader adoption.

\subsection{Authoring Data Videos}
While data videos are a powerful medium for storytelling, creating them requires significant expertise in data analysis and video crafting. General-purpose video editing tools like Adobe Premiere Pro~\cite{premiere} allow users to manipulate visual elements through animation keyframing and presets, but they often involve extensive manual effort and a steep learning curve. 
To lower the entry barrier, researchers have developed systems to simplify the creation process.

Early systems like DataClips~\cite{DBLP:journals/tvcg/AminiRLMI17} introduced template-based approaches, enabling semi-automatic data animation by reusing patterns from existing examples. Building on this foundation, tools such as WonderFlow~\cite{DBLP:journals/corr/abs-2308-04040} and InfoMotion~\cite{DBLP:journals/cgf/WangGHCZZ21} began automating more complex aspects of storytelling, including synchronizing animations with audio narrations and applying animations to static infographics by analyzing their structures. Systems like Gemini~\cite{DBLP:journals/tvcg/KimH21} and its extension Gemini2~\cite{DBLP:conf/visualization/KimH21} focused on refining the transition process by recommending staged animations, offering designers guidance for creating smooth and semantic transitions. 
Meanwhile, tools like AutoClips~\cite{DBLP:journals/cgf/ShiSXLGC21} and Roslingifier~\cite{DBLP:journals/tvcg/ShinKHXWKKE23} emphasized automation and customization, with the former generating videos from data facts and the latter supporting semi-automated narratives for scatterplots. 
Recent advances, such as Live Charts~\cite{DBLP:journals/corr/abs-2309-02967} and Data Player~\cite{DBLP:journals/tvcg/ShenZZW24}, further integrate intelligent automation, using large language models to link narration and visuals, while Data Playwright~\cite{DBLP:journals/corr/abs-2410-03093} pushes the boundary by introducing natural language-driven video synthesis through annotated narration. 
Together, these tools represent a progression from template-based semi-automation to intelligent systems that combine automation, user input, and natural language interaction, making data video authoring increasingly accessible and efficient.

\SB{} differs from these systems in two key aspects: (1) it focuses on authoring augmented sports videos with embedded data visualizations rather than general-purpose data videos; and (2) it emphasizes real-world deployment by collecting real user data to refine its usability and functionality. This approach bridges the gap between academic research and practical application, addressing the unique challenges of sports video storytelling in authentic settings.

\section{\SB{} Design}
\subsection{Gaps in Sports Video Storytelling
}
To design an easy-to-use interactive sports video storytelling tool, we conducted over 50 user interviews with diverse sports roles with the need to share insights through game videos. These roles cover coaches, athletes, trainers, parents, marketing specialists, media producers, self media content creators, and sports fans. 

While their needs for creating and sharing sports videos differ, they all aim to convey their insights in videos to effectively tell stories and engage their audience. The majority of these sports domain users found high entry barriers of existing video editing tools due to lack of proper skills and time. Three gaps were identified: 

\textbf{1. Popular video editing tools have steep learning curve and lack features tailored to sports videos.} Media professionals, such as YouTubers and media specialists, commonly use video editing tools like Adobe Premiere Pro~\cite{premiere}, iMovie~\cite{imovie}, or web platforms like CapCut~\cite{capcut}. However, other sports roles often lack the skills to use these tools, facing a steep learning curve. Even for media professionals, these tools lack sports-specific features, requiring manual customization, such as player tracking or tactical annotations, which is both time-consuming and challenging. For non-video professionals, creating effective sports highlights is often out of reach, despite the significant impact such highlights could have on communication and profile building. This gap makes it difficult for many in the sports industry to leverage video storytelling effectively.
\textbf{Design Implication:} An effective sports highlight creation tool should provide an easy workflow with dedicated sports-specific features.

\textbf{2. Limited access to sports-specific video analysis and creation tools.} 
While tools like Hudl~\cite{hudl} and WSC Sports~\cite{wscsports} offer video clipping and analysis for sports, they are often expensive and designed for team-level access, making them inaccessible to individual coaches, athletes, and creators.
Additionally, these tools are primarily geared toward game analysis or recaps rather than creating sports highlights that effectively convey insights. 
Solutions for sports broadcasters, like Piero~\cite{piero} and Viz Libero~\cite{viz-libero}, allow for the easy addition of sports graphics and annotations, but are typically costly and require complex setups within a media production pipeline. 
As a result, most individuals in sports ecosystem lack access to these specialized tools, limiting their ability to leverage video for their storytelling. 
\textbf{Design Implication:}
A sports video highlight tool tailored to individual users should be accessible, flexible, and function independently without the need for complex team setups.

\textbf{3. Disruptive video creation workflow is distracting and time consuming.} 
Videos play a crucial role in the workflows of many sports professionals, such as coaching and performance analysis. While they often have systems in place to manage game clips, creating highlight videos with personal insights typically requires manual switching between multiple tools and labor-intensive editing to produce meaningful content.
This additional burden diverts professionals away from their primary tasks. Consequently, the effort required to create effective highlights often outweighs the benefits, discouraging regular use of video editing tools to communicate insights.
\textbf{Design Implication:} An effective sports highlight creation tool should seamlessly integrate video management and editing workflows to streamline the creation process.

\subsection{Design Goals}
As \SB{} targets sports domain experts and fans to allow them to convey personal insights through video storytelling effectively, to address the aforementioned gaps,
we aim to design an interactive system that supports easy highlight creation tailored to sports videos and individuals with four design goals below:

\textbf{G1: Intuitive sports features and interactions.} Video editing softwares can be overwhelming for beginners due to unintuitive interfaces and complex steps that distract from the communication intent. For example, adding an animated arrow requires navigating menus, adjusting settings in separate panels, and applying animation in a disconnected interface. This process is far less intuitive than how sports experts simply draw arrows on a tactical board to convey their insights.
To make video editing more accessible for sports experts, tools should provide easily discoverable features and context-embedded interactions that closely align with user intent to lower the entry barriers.

\textbf{G2: Object level visualizations.} Naturally, storytellers describe game insights from the human perspective and apply visual aid to bring attention to the players and their actions, such as highlighting the shooter with a spotlight. Traditional video editing method, which operates at the graphic level, require creators to move a spotlight layer frame-by-frame and manually follow the player position, which is time consuming and tedious. We aim to eliminate such need by automatically attaching visualization to the object based on player tracking and segmentation, allowing easy addition of player-specific visualizations, which aligns better with users' mental model.

\textbf{G3: Seamless integration of visuals and narratives.} Sports insights are typically shared through visual highlights, tactical drawings, and narratives alongside the game footage. Traditionally, coaches and broadcasters rely on voiceovers and pointers or separate tactical boards to explain key moments. However, insights tied to specific players or actions on the court often require direct integration into the video for better clarity and engagement.
To effectively communicate insights in sports highlights, tools must provide features to integrate visual aids and narratives into game footage easily.


\textbf{G4: Streamlined end-to-end video creation process.}
To empower users to leverage videos for sharing their insights effectively, it is crucial to support their workflow with the fewest manual steps possible. From importing video clips and editing to sharing, these steps should have minimal context switching to allow an intuitive and smooth user experience.


\begin{figure}[t!]
    \centering
    \includegraphics[width=\linewidth]{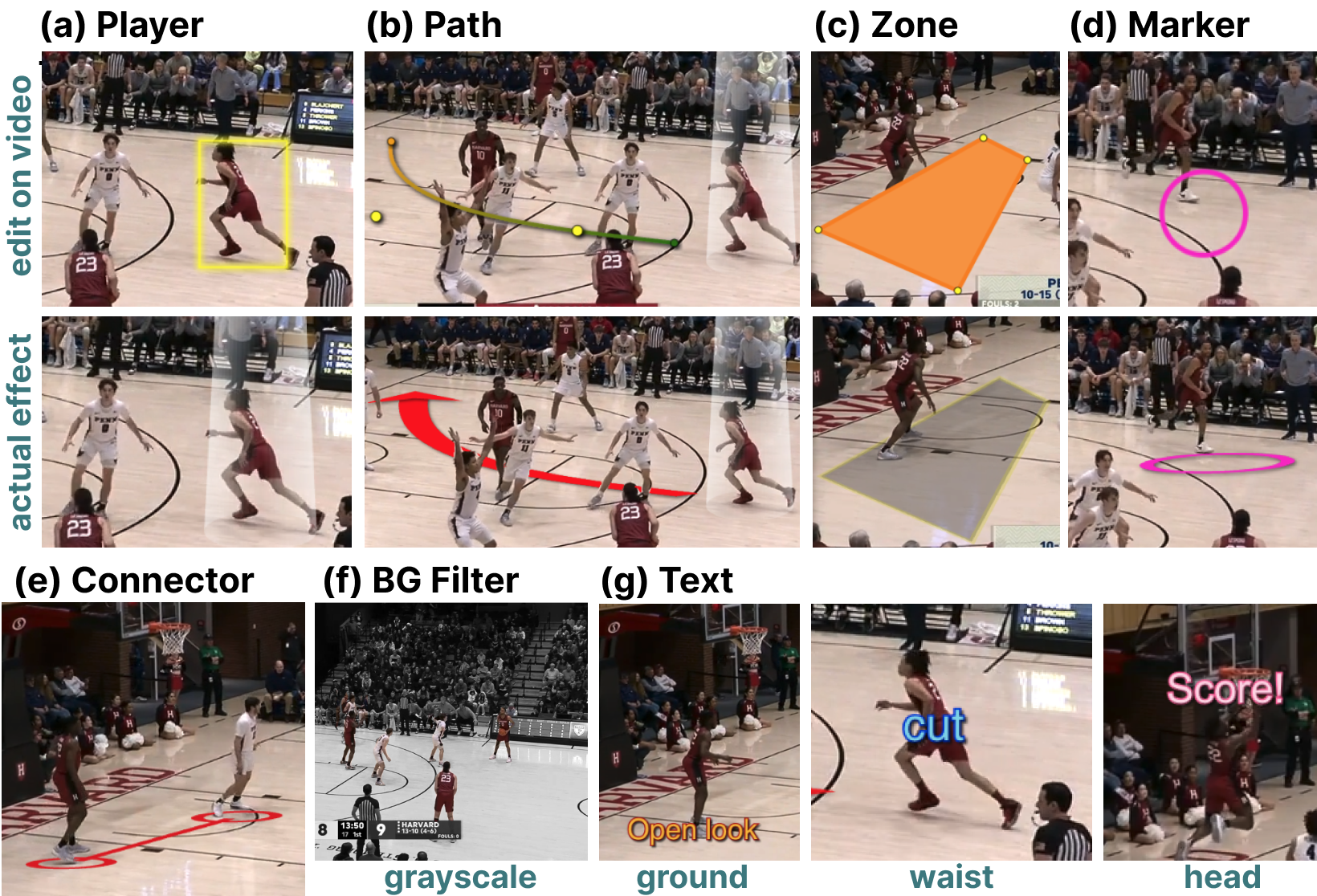}
    \vspace{-5mm}
    \caption{Nine visualization features are provided under \Highlight{} Tab. Users can directly interact with the video to apply effects, such as clicking on a player (a) or drawing on canvas (b-d). These features support communicating sports insights, including highlighting players (a)(e), illustrating tactics (b-d)(f), and annotating actions (g).} 
    \label{fig:vis-feature}
    \vspace{-3mm}
\end{figure}

\subsection{Visualization Design}
\label{sec:vis-feature}

Based on our user interviews and literature reviews, sports insights in relation to game performance can be categorized into three main types: 
\textbf{1) Emphasize key players}, where the storyteller wants to \textit{bring attention to a player's motion or position};
\textbf{2) Illustrate tactical strategy}, where the user wants to \textit{showcase the detail breakdown of action sequences}, often involving the spatial relationship among players/teams and temporal movement;
\textbf{3) Associate insights to the performance}, where the user wants to \textit{directly annotate decisions or outcomes} to link their insights to the game actions.

To support each insight type with video storytelling, we design nine visualization features under three categories, including player, tactic, and action, as shown in Fig.~\ref{fig:teaser} (a2):

\begin{itemize}[leftmargin=*]
    \item \textbf{Player highlights.} To bring attention to key players, \Circle{} and \Spotlight{} effects apply visual overlay on individual players, while \Connector{} links two or more players to emphasize their spatial relations. For example, \Circle{} and \Spotlight{} are frequently used to focus on key players involved in the play or have outstanding performance, such as the scoring player. \Connector{} is most popular in team sports like soccer, to bring attention to certain attack or defense formation (Fig.~\ref{fig:vis-feature} (e)).

    \item \textbf{Tactic highlights.} To provide deeper insights in sports highlights, it is essential to support tactical drawing, a common practice on sketch boards, directly within the video. As shown in Fig.~\ref{fig:vis-feature} (b)-(d), \Path{}, \Zone{} and \Marker{} allows users to annotate player trajectories, areas, and positions using sports-specific symbols, such as ``O" for offense players and ``X" for defense. To mimic the intuitive interaction of a tactic board, users can draw symbols and lines directly on the video. To prevent visual clutter from excessive annotations,
\BGFilter{} applies a background color to the video, bringing players and tactic visualizations to the forefront with clear contrast (Fig.~\ref{fig:vis-feature} (f)). Additional video effects, such as \textit{Freeze} frame and speed adjustments, serve as complementary tools to further emphasize tactical insights.

    \item \textbf{Action highlights.} 
To connect the insights to game performance, users can directly highlight actions in the video with their annotation. \Zoom{} feature allows direct change of video view to zoom in and follow a selected player. 
\Text{} feature support both adding text at a fix position or track a player, allowing users to directly label actions or insights in-related to the game action, such as a player's move or outcome. Three text label placement options, head, waist, and ground, also provide more flexibility for annotating different insights, such as showing the observation ``Open look'' below the player, the action ``Cut'' on the waist, and the outcome ``Score!'' above the player's head (Fig.~\ref{fig:vis-feature} (g)).
\end{itemize}

Overall, these sports-specific visualizations are designed to address the most common insights related to players, tactics, and actions, drawing inspiration from basketball, soccer, and feedback from coaches and creators. 
In addition, with its easily adjustable timeline tracks, \SB{} highlights features empower users to compose and customize visualization features flexibly, allowing them to craft their own communication toolkit and styles to effectively convey their insights.

\begin{figure*}[ht!]
    \centering
    \includegraphics[width=\linewidth]{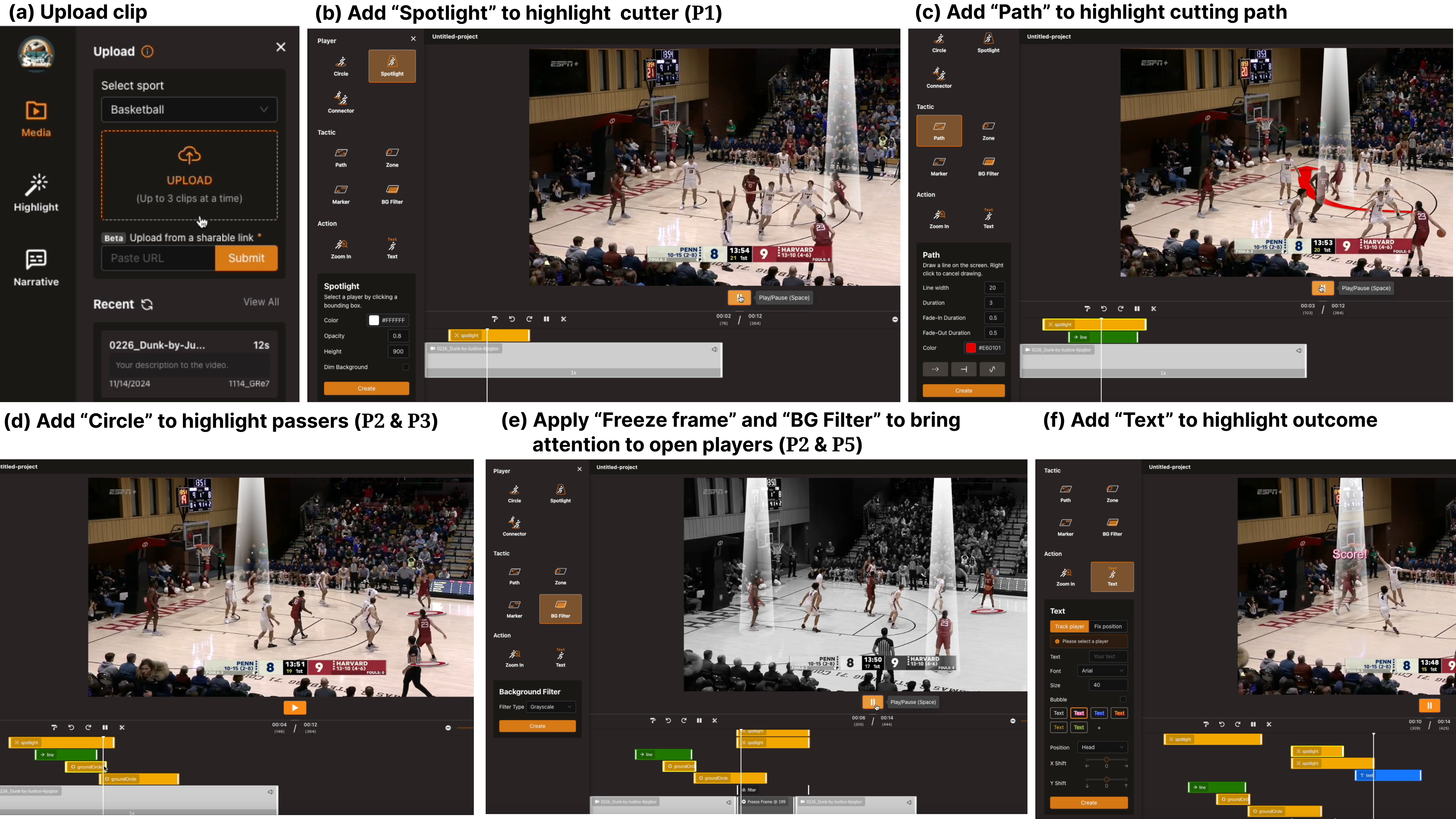}
    \vspace{-6mm}
    \caption{A demonstration of a coach adding insights with visualization features on \SB{}: starting with uploading a clip (a), highlighting player movements and tactics (b-d), and annotating insights with a video freeze frame, background filter, and text labels (e-f).
    } 
    \label{fig:user_journey}
    \vspace{-3mm}
\end{figure*}

\subsection{User journey}
To illustrate the practical application of \SB{}, we present a coach's user journey in creating an engaging sports highlight to explain a tactical breakdown to the team, as shown in Fig.~\ref{fig:user_journey}.

Coach Mike uploads a game clip to \SB{} under \Media{} (Fig.~\ref{fig:user_journey}a). In the clip, his team executes smooth passes and cuts, resulting in an easy dunk by the center. Mike wants to highlight the timing of player actions and provide positive feedback to the team .
He starts by opening \Highlight{} and selecting the \Spotlight{} feature (Fig.~\ref{fig:user_journey}b). Clicking on player \texttt{P1}, who makes a pass to \texttt{P2} and cuts to the opposite corner, Mike easily applies a spotlight effect that automatically tracks \texttt{P1}. He adjusts the spotlight's duration in the timeline to bring attention on the cutter's movement. 
Next, he uses the \Path{} tool to draw the cutting path, illustrating the correct route (Fig.~\ref{fig:user_journey}c).
To emphasize follow-up actions by the other players, he adds \Circle{} effects to highlight \texttt{P2} and \texttt{P3} as they pass the ball (Fig.~\ref{fig:user_journey}d). 
When \texttt{P3} gains possession, he highlights two options: passing back to \texttt{P2} for a three-point shot or passing to \texttt{P5} for a dunk. To emphasize these options, Mike freezes the frame by clicking on the freeze icon, applies a \BGFilter{} to grey out the background, and uses the \Spotlight{} tool to bring focus on \texttt{P2} and \texttt{P5} (Fig.~\ref{fig:user_journey}e). After the pass, he annotates \texttt{P5} with a \Text{} effect, adding the label ``Score!'' to highlight the successful execution (Fig.~\ref{fig:user_journey}f).

Finally, Mike switches to \Narrative{} to add captions explaining the tactic. He finishes the highlight and clicks on \textit{Export} to easily share this game highlight link with his team. 
The entire creation process took Mike less than 5 minutes.
\section{System Design and Implementation}

We developed \SB{},  an AI-powered platform for creating sports video highlights, featuring intuitive visualization tools designed for effective insight communication.
\SB{} system comprises three core components:
\textbf{1) AI-powered Video Processing Pipeline.} A server dedicated to processing all uploaded videos using advanced computer vision techniques.
\textbf{2) Web-based Video Editor.} A React.js-based interface where users can perform video editing and explore visualizations.
\textbf{3) Cloud-Based Distributed Server.} A Python Flask server that facilitates seamless data communication between the user interface and the backend.

In this section, we detail the video processing pipeline and the methods used to present visualizations in the web app. We also discuss how we have streamlined the creation process to enhance user experience, and how AI is leveraged to facilitate content creation. Lastly, we address the current limitations of our system.

\subsection{Video Processing Pipeline}
\label{sec:video-processing}
We use a video processing pipeline to pre-process sports videos for editing. 
The pipeline includes three parallel branches: 

\begin{itemize}[leftmargin=*, topsep=1pt]
    \item \textbf{Player Detection.}
This branch performs two tasks: (1) player tracking and (2) pose estimation. For player tracking, we employ MixSort~\cite{cui2023sportsmotlargemultiobjecttracking}, a customized tracker optimized for sports datasets. For pose estimation, we use MMPose~\cite{mmpose2020}, a top-down approach that utilizes bounding boxes from player tracking to predict key points for each player. Together, these methods ensure accurate and efficient tracking of player movements and poses.

    \item \textbf{Foreground-Background Segmentation.}
This branch uses Mask R-CNN~\cite{he2018maskrcnn} to segment human figures (foreground) from the court (background). The resulting segmentation masks are exported in MP4 format for use in the web app. 

    \item \textbf{AI Captioning.}
The AI Captioning branch leverages GPT-4o, a large language model (LLM), to process video content and generate detailed descriptions. Frames are sampled at a rate of one frame every 30, encoded, and then processed by GPT-4o to extract relevant insights, such as team and player details. The insights are formatted into a JSON file with second-by-second annotations and subsequently refined by another LLM to synthesize captions tailored to user needs.
\end{itemize}

The pipeline's parallel architecture operates asynchronously, ensuring efficient video processing. Each branch demonstrates robust performance, making the system well-suited for production environments. \revision{For detailed accuracy and performance evaluations of these methods, readers are referred to the original works~\cite{cui2023sportsmotlargemultiobjecttracking, mmpose2020, he2018maskrcnn}. Our primary contribution is integrating these methods into a unified visualization system and applying them in real-world, user-centered scenarios.}

\subsection{Web-based Video Editor}

\subsubsection{User Interfaces}
\SB{} interface, as shown in Fig.~\ref{fig:teaser}, consists of 
(a) a feature menu with a collapsible panel for the selected feature list, 
(b) a video canvas featuring an interactive video view and a timeline, and 
(c) a top menu bar for video export and account setting. 
To support an end-to-end highlight creation workflow, users can upload their clips, edit, and share the created sports highlight all in one place (\textbf{G4}). 

\begin{figure}[t!]
    \centering
    \includegraphics[width=\linewidth]{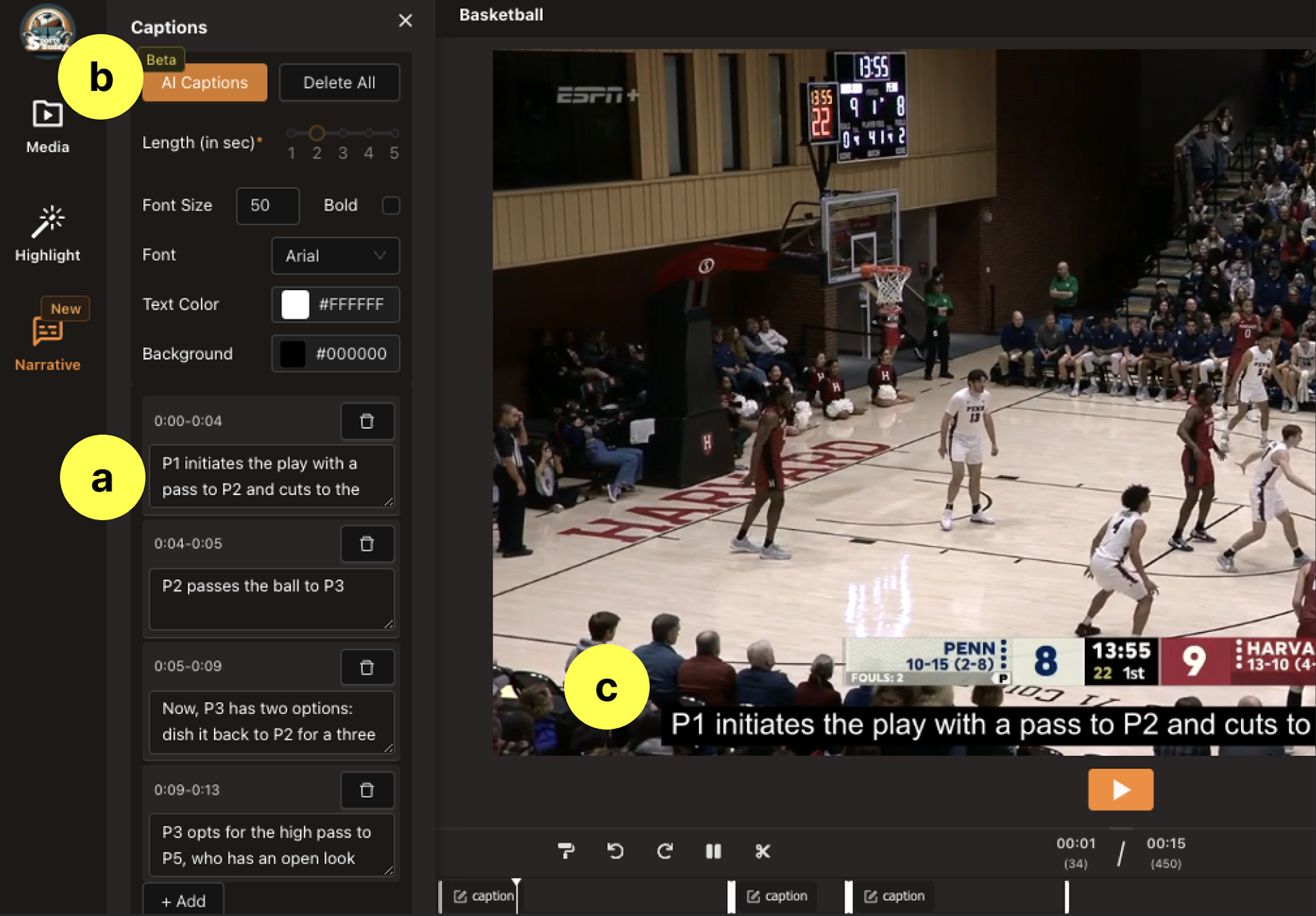}
     \vspace{-6mm}
    \caption{\Narrative{} Tab provides captioning features. Users can (a) input captions manually, (b) create AI-generated captions, and (c) see instant caption update in the video.} 
    \label{fig:narrative}
    \vspace{-3mm}
\end{figure}

\vspace{1mm}
\noindent\textbf{Feature Menu.}
Three key feature categories are provided on the menu sidebar (a1), including \Media{}, \Highlight{}, and \Narrative{}. Users can click on each tab to see the respective feature list (a2):

\begin{itemize}[leftmargin=*]
    \item \emph{Media Tab.} A user can upload and manage video clips under \Media{} tab, as shown in Fig.~\ref{fig:user_journey} (a). They can select sports types from five supported sports, including basketball, soccer, volleyball, lacrosse, and tennis, and update clip through drag-and-drop, or paste a URL of an online video link. They can retrieve  clips by selecting from the recent upload list or clicking ``View All'' to search for clips in the media library. 
    
    \item  \emph{Highlight Tab.} Once a clip is loaded into the video canvas, the user can select from various sports highlight effects to turn their insights into engaging visualizations (\textbf{G1}). The highlight features are grouped by the type of sports insights, including players, tactics, or actions, as shown in Fig.~\ref{fig:teaser} (a2). When selecting a highlight feature, the user will see the corresponding feature panel (Fig.~\ref{fig:teaser} (a3)) to create the visualization or customize the color, shape or other settings. After the visualization is created, each element is mapped to an interactive track shown on video timeline (Fig.~\ref{fig:teaser} (b3)). These effect tracks are encoded with colors, orange for player features, green for tactic features, and blue for the action features.
    Detailed visualization design considerations are described in Sec~\ref{sec:vis-feature}.
    
    \item  \emph{Narrative Tab.} The user can add captions to the video under \Narrative{}, as shown in Fig.~\ref{fig:narrative}. Captions can be typed directly, with each line mapped to a corresponding subtitle track on the video timeline. The user can customize the font style and color, with real-time updates displayed on the video. Additionally, the user can select \textit{AI Captions} feature to automatically generate subtitles, with options to create sentences from 1- to 5-second video segments.
The user can interactively adjust the duration of each subtitle track on the video timeline. 
\end{itemize}

\vspace{1mm}
\noindent\textbf{Video Canvas.}
The video canvas consists of a video preview (b1), an editing toolbar (b2), and a video timeline (b3):

\begin{itemize}[leftmargin=*]
    \item \emph{Video Preview.} Apart from playing the video preview, the user can directly interact with the video to apply visualization effects (\textbf{G2}), as depicted in Fig.~\ref{fig:vis-feature} (a)-(d). For player and action features, the user clicks on the player in the video to create visualizations or zoom in to the selected player. For tactic features, the user draws path, zone, or marker position on the canvas. They can also preview the video in full screen mode. 

    \item \emph{Editing Toolbar.} Five common editing features are supported, including \textit{Reset}, \textit{Undo}, \textit{Redo}, \textit{Freeze}, and \textit{Split}. \textit{Freeze} allows users to add a static frame to the selected video frame, commonly used in explaining tactics in game breakdown. \textit{Split} separates the video track into two, allowing duplication or adjusting speed for individual tracks in the video timeline. Each editing tool has a shortcut key to support fast video editing. 

    \item \emph{Video Timeline.} Each track on the video timeline represent an added highlight or narrative effect, with the original video track at the bottom (Fig.~\ref{fig:teaser} (b3)). Users can change the duration and timing of the effects by moving the tracks or deleting them, providing a comprehensive overview of all visuals and narratives at one place (\textbf{G3}).
    On top of it, right click on the video track will show a track submenu for users to \textit{Mute}, \textit{Duplicate}, or change \textit{Speed} of the video easily, as shown in Fig.~\ref{fig:teaser} (b4).
\end{itemize}


\vspace{1mm}
\noindent\textbf{Top Menu.}
The top menu bar provides convenient access to export videos, open tutorial, and additional settings.
\begin{itemize}[leftmargin=*]
    \item \emph{Export.} To streamline the video creation process, users can export their highlights directly as a video file, a shareable link, or post them to popular social media platforms for sports content, like X, Facebook, or Reddit, as shown in Fig.~\ref{fig:teaser} (c1).
    If specified, exported videos can be stored in publicly accessible cloud storage in combination with social media friendly HTML tags. The two setups making it simpler and faster to direct share to social media platforms such as Facebook and X, streamlining content distribution for users.

    \item \emph{Additional Settings.}  To ensure future scalability, additional setting options are available.
Users can manage their account information and language preferences. Currently, \SB{} supports English, Simplified, and Traditional Chinese, with the flexibility to expand to more languages for the global sports community.
\end{itemize}


\subsubsection{Canvas-based Video Rendering}
Visual effects in the \SB{} web app are rendered using an HTML canvas.
To enable object-level control over visualizations (\textbf{G2}), 
we introduce the concept of a \textit{render object}.
A render object is a structured data entity that encapsulates all the necessary information for rendering a specific visual effect.
Each render object includes key attributes for UI and rendering, such as its \emph{type}, \emph{start frame}, \emph{end frame}, and \emph{effect parameters}.
\SB{} supports a total of 13 render objects, including 9 visualizations described in Sec.\ref{sec:vis-feature}, 
as well as \textit{Caption}, \textit{Freeze Frame}, \textit{Background}, and \textit{Foreground}.


\setlength{\itemsep}{0pt} 
\setlength{\parskip}{0pt} 
\setlength{\topsep}{0pt}  

\vspace{1mm}
\noindent\textbf{Rendering Strategy.}
Most render objects represent individual visual effects, positioned within the canvas coordinate system with their shapes and positions calculated in real time. Different types of render objects employ distinct rendering strategies:
\begin{itemize}[leftmargin=*]
    \item \emph{Player highlights}. 
    Render objects like \textit{Circle} and \textit{Spotlight} are dynamically rendered at the coordinates of the selected player's bounding box in the current frame, allowing them to move with the player.

    \item \emph{Tactic highlights}. 
    Render objects such as \textit{Path} and \textit{Zone} are rendered at fixed positions based on user-provided coordinates to highlight tactical elements.

    \item \emph{Action highlights}. Render objects for action highlights modify the canvas attributes or states rather than representing direct visual effects. Key render objects include: \emph{BG Filter} -- Adds a filter effect to the canvas; \Zoom{} -- Applies a scaling transformation to the canvas; and \emph{Freeze Frame} -- Provides a static image source to represent a frozen moment in the video.
    
\end{itemize}

Additionally, for all visual effects that need to appear on the ground, such as \textit{Circle} and \textit{Area}, a transformation matrix is applied to the original coordinates. 
This transformation creates a realistic three dimensional effect that aligns the visualizations with the court's perspective.

\vspace{1mm}
\noindent\textbf{Rendering Order.} 
To embed the visualizations into the scene, we place visualizations in two key locations:
\begin{itemize} [leftmargin=*]
    \item \emph{Between the Background and Foreground.} Suitable for effects interacting with the court, such as areas and markers.
    \item \emph{Above the Foreground.} Designed for overlays that highlight players or events, such as \textit{Spotlight} and \textit{Text} annotations.
    
\end{itemize}

To ensure proper stacking and organization, we introduce a \textit{layer} concept. Each rendering order corresponds to a specific layer, helping to manage the visual hierarchy. Render objects are processed and displayed at their assigned layer during each frame. Smaller numbers indicate layers that are rendered first.

\vspace{1mm}
\noindent\textbf{Rendering Efficiency.}
For all the render objects rendering on canvas, we are able to control the overall rendering time within 10 milliseconds for each frame. This means we support a 60 FPS video playing smoothly.

\subsection{Cloud-Based Distributed Server}
To efficiently handle user requests and enable scalable video processing, we implement a distributed server architecture. At the core of the system is a Python Flask server, which facilitates communication between the frontend and backend. User requests are logged into a distributed message queue system, such as Kafka, ensuring reliable and asynchronous task management.
Multiple GPU server nodes are deployed to pull requests from the queue and process the video tasks. Once processing is complete, the results are stored in a centralized database, which is accessible to the Flask server for seamless integration with the frontend.

The entire architecture is deployed on the Azure cloud platform, leveraging elastic scaling to dynamically allocate resources based on user demand. This setup ensures robust performance, high availability, and efficient handling of variable workloads.

\subsection{System Limitation}
The current system has limitations affecting both the web app and the video processing pipeline.

\vspace{1mm}
\noindent\textbf{Video Processing.}
The primary limitation of the video processing pipeline lies in the player tracking algorithm. It can fail under common conditions such as occlusion, changes in player's gesture, or player disappearance and reappearance—issues frequently encountered in sports videos. To address these challenges, improvements to the tracking algorithm are required, along with fine-tuning it on our specific dataset for more robust performance.

\vspace{1mm}
\noindent\textbf{Visualization Rendering.}
The current implementation lacks parameters of the camera, which impacts certain visual effects, such as \textit{Path} and \textit{Zone} that require a realistic 3D perspective. Currently, we apply a fixed transformation matrix across all scenarios, which can result in suboptimal visual quality in specific cases. Additionally, as the camera angle changes and the court shifts, the visualizations do not dynamically move with the court as the render objects are positioned in the static canvas coordinate system. Without proper coordinate mapping to the actual court, there could potentially be misalignment between visualizations and the court’s position.

\vspace{1mm}
\noindent\textbf{Web App Performance.}
Since the application runs in a web browser, its performance is inherently tied to browser memory usage. For instance, rendering a 15-second, 1080p video clip typically uses around 200MB of memory, which is currently acceptable. However, supporting longer or higher-resolution videos will require us to implement dynamic loading and compression mechanisms to maintain optimal performance.

\section{Real-world system evaluation}
We have made \SB{} publicly available online\footnotemark[1] since August 2024. 
After three months of deployment, \SB{} has attracted 163 registered users. We collect user background and feedback to evaluate the usefulness and benefits of \SB{}. During these three months, we have also made feature and user experience improvements based on user feedback. 

In this section, we analyze user feedback and insights that inform critical system improvements, followed by two case studies on the usefulness of \SB{} in supporting athletic communication and game analysis and video storytelling on social media.

\subsection{User Feedback}


Of the 163 users, 66 completed a demographic questionnaire. Among them, 38\% identified as coaches, 20\% as content creators, 14\% as athletes,  12\% as parents,  7\% as fans, 2\% marketing specialists, 
and 7\% as others. In terms of age distribution, 40\% of users were aged 20–30, followed by 29\% aged 30–40, 24\% aged over 40, and 7\% aged below 20.
Based on user feedback collected over three months, we found that \SB{} enhances user workflow and video quality, and engagement with efficiency and creativity. Further, we identified the system's strengths and areas for improvement, leading to several enhancements implemented during this period.

\subsubsection{Usefulness \& Impact}



Above all, \SB{} significantly enhances video creation efficiency. Many users reported substantial reductions in editing time, with one stating that tasks previously taking over 30 minutes could now be completed in less than 10 minutes. The automation of complex processes, including tracking player location with player features and visual annotations with \Text{} features, was highlighted as a key time-saver, especially for less experienced users like interns and youth coaches.

\revision{Quantitative insights gathered during this period provide additional evidence of the system's effectiveness. A total of 1,021 videos were successfully uploaded, with an upload success rate of 90.8\%, and 814 highlights were exported, achieving an export success rate of 87.9\%, with no major disruption reported in the user feedback. These metrics reflect the reliability of the video processing pipeline and its ability to handle diverse video inputs and produce high-quality outputs efficiently. Follow-up interviews with users further validated these findings, with users reporting seamless functionality in AI-enabled features like player tracking and pose estimation.}


The most popular features were \Spotlight{}, \Circle{}, and \Text{} annotations, which users praised for enhancing their storytelling with clear visual guidance and embedded narratives. Some users also explored creative usage of less common features, such as \Connector{} and \BGFilter{}. For instance, a content creator used \Connector{}, typically applied in soccer to highlight team formations, to illustrate defensive setups in basketball. Similarly, \BGFilter{}, perceived as a novel visualization effect for sports highlights, enabled users to create strong visual contrasts, further directing the audience's attention to key player actions and moments.

Overall, users consistently acknowledged \SB{}'s ability to enhance audience engagement through visually compelling content. 
Coaches particularly appreciated \SB{} for its precision in identifying areas where players need improvement, enabling clearer and more actionable feedback for athletes. This approach aligns with their principle of \textit{``Show, don't tell},'' enabling effective use of visual demonstration over verbal explanation. 
Social media managers and influencers valued the tool’s ability to create engaging highlight clips quickly, which helped increase their reach and follower engagement.
In addition, parents of youth athletes highlighted \SB{}'s potential for creating high-quality player highlights, which are essential for recruitment, scholarships, and applications to collegiate or youth teams. 

\subsubsection{Insights for System Improvements}

Users feedback has also highlighted areas for improvement in functionality and feature design, leading to improvements in supporting more streamlined user workflow and scalable function extension.

\textbf{1. Enhanced Video Processing Workflow}. While users found \SB{}'s interface intuitive, several raised concerns about long processing times and the lack of batch processing options. These limitations reduced efficiency for high-volume workflows, such as creating multiple highlight reels or analyzing full-game footage. 
In response, users recommended adding batch upload and processing capabilities to streamline their workflows.

To address processing time concerns, we optimized backend systems with multiple thread processing and better GPUs to improve processing speed by 2 to 3 times, and refined the user interface to specify the progress made, such as showing the current video processing steps and estimated remaining time, to allow a smoother and more transparent user experience. 
Additionally, we introduced a batch upload feature, allowing users to upload multiple clips in one session. This enhancement allows users to have a coherent workflow without interruption.

\textbf{2. Customization and Automation}.
Most users found the sports-specific visualization tools intuitive and effective, but some requested greater flexibility. For example, the \Zoom{} effect lacked dynamic keyframe control, limiting its adaptability. 
On the other hand, several users expressed interest in features that go beyond video editing to deliver deeper sports insights, such as live tactical analysis and automatic detection of key plays.

To balance diverse user needs with our design goal of simplifying highlight creation for sports professionals, we designed \SB{} to incorporate both customization and automation. Beyond the basic sports features in \Highlight{}, feature panels (Fig.~\ref{fig:teaser}(a3)) are designed to accommodate customization options such as styles and templates, allowing users tailor their content to specific needs. 
Additionally, on top of the typical captioning tool under \Narrative{},
we introduced an AI captioning feature, described in Sec.~\ref{sec:video-processing}, to deliver alternative insights through intuitive interactions.  While this represents just one approach to AI automation in sports highlight creation, our interactive design allows users to easily refine and adjust AI-generated content. This foundation is built with scalability in mind, paving the way for future AI model enhancements and fostering seamless human-AI collaboration.

\subsection{Case Study}
To understand how \SB{} supports video storytelling in real-world setting, we conducted two case studies. In the first study, we collaborated with Harvard Athletics Department to enhance the diversity of social media content for basketball and soccer teams. In the second study, we collaborated with basketball influencers on Youtube and Instagram to enhance their content quality and simplify the video editing process. In both cases, we discuss the types of video stories they made, and highlight the unique values \SB{} brought to their workflows and outcomes.

\begin{figure}[t!]
    \centering
    \includegraphics[width=\linewidth]{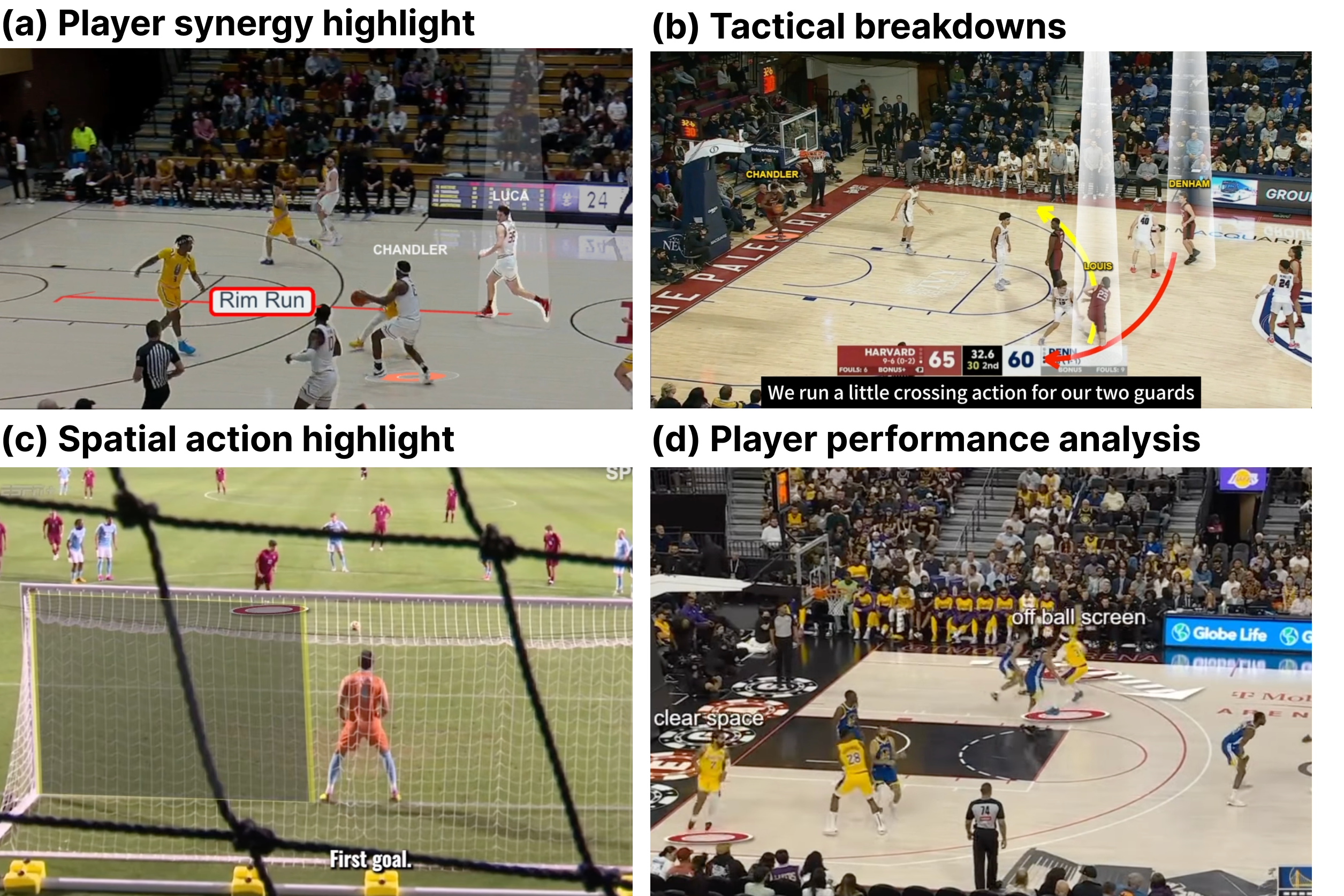}
    \vspace{-6mm}
    \caption{\SB{} enhances the effectiveness and diversity of video creation for athletic communication, game analysis, and storytelling, supporting use cases such as highlighting (a) player synergy, (b) tactics, (c) spatial actions, and (d) player performance.} 
    \label{fig:case_study}
    \vspace{-3mm}
\end{figure}

\subsubsection{Athletic Communication}
Amateur and collegiate teams often lack the advanced tools and resources needed to produce high-quality highlight reels, such as those seen in professional leagues like NBA. \SB{} presents a new opportunity to enrich game highlights with more personal insights that enhances team visibility and fan engagement on social media.
Using \SB{}, marketing professionals and coaches at Harvard Athletics created innovative game highlights with engaging visuals and insightful narratives that were previously unavailable. Seven basketball highlights and seven soccer highlights were produced and shared on Instagram. Theses 14 videos have collectively attracted over 150 million views. 
We gathered user feedback and identified three unique types of insights made with \SB{} highlights.

\textbf{1. Highlight player synergy}. Before using \SB{}, teams often share the best game moments to showcase team success, such as a winning shot, often focusing on a single player with multiple camera angles and close-ups. While this content is highly engaging due to its dynamic motion,  it often overlooks the interplay among players critical to team success. With \SB{}, teams can draw attention to the synergy between key players. For example, as shown in Fig.~\ref{fig:case_study}a, a smooth rim run paired with a precise pass between two players was highlighted, showcasing a successful transition play driven by teamwork.

\textbf{2. Tactical breakdowns.} \SB{} also enables teams to seamlessly convey coaching strategies in game highlights, enhancing the depth of their storytelling. In the example shown in Fig.~\ref{fig:case_study}b, a coach's verbal commentary was turned into clear, integrated visual highlights. Traditionally, such insights were conveyed verbally or as text alongside raw game footage, often making them difficult to comprehend on social media. With \SB{}, visual highlights and narratives work together effectively, allowing audiences to grasp the complexity of the strategy without feeling overwhelmed.

\textbf{3. Emphasize spatial actions}. The third type of insights target emphasizing the spatial elements of gameplay. For example, in a soccer scoring highlight (Fig.~\ref{fig:case_study}c), a \Zone{} visualization was used to highlight the scoring area on the net, alongside a \Circle{} marking the kicker. In a basketball clip, the distance between the shooter and the hoop was visually emphasized. These visual enhancements draw attention to the spatial attributes of the action, helping audiences appreciate the precision and difficulty involved.

Overall, the feedback from Harvard Athletics was overwhelmingly positive. Collaborators quickly learned to use the tool and applied its features creatively to support their storytelling needs. The innovative ways in which each feature was utilized highlight \SB{}'s effectiveness in reducing operational burdens while empowering users to focus on the creative process.

\subsubsection{Game Analysis and Storytelling}

Individual sports content creators are often sports fans with varying levels of video or image creation skills. Due to limited time, skills and access to professional tools, they face significant challenges in producing high-quality videos to deliver their unique insights. 

To explore how \SB{} can empower individuals in creating engaging highlights for game analysis and storytelling, we collaborated with three basketball influencers, each offering unique perspectives: \textit{HungKu}, a popular basketball YouTuber with over 100K followers covering NBA news~\cite{hungku}; \textit{Hinbasket}, a basketball-focused Instagram account with 12K followers specializing in game analysis and team tactics~\cite{hinbasket}; and \textit{JNC\_bball}, an Instagram account with 7K followers that analyzes player performance~\cite{jnc}.
With distinct focuses in their highlights, each creator found \SB{} supported their workflow and enhanced their content in unique ways. 

\textbf{1. Enhance video quality and engagement.} 
With the goal to delivers timely NBA news through long-format YouTube videos, HungKu integrates multiple short game clips to support his storytelling. In longer videos, smooth transitions and engaging visuals are essential for maintaining audience attention. With\SB{}, HungKu can create high-quality highlight clips with features that spotlight players and guide audience attention. He reported that \SB{} seamlessly integrates into his workflow, enhancing video quality and effectiveness with minimal effort.

\textbf{2. Create engaging tactical short reel.} Hinbasket creates short highlight to cover key moments from NBA games. He focuses on delivering his insights on team tactics and game performance, often requiring annotation of specific players or movement. Previously, this process was highly time-consuming. With \SB{}, he leverage AI to easily track players and add tactical elements like arrows and labels, making his insights stand out. Overall, \SB{} has significantly reduced his editing time and improved clarity and engagement of his highlight reels.

\textbf{3. Unlock new formats for player analysis.} Before using \SB{}, JNC\_bball primarily shared player performance insights through static images and text, which had already garnered them a loyal following due to their thoughtful and fresh perspectives. However, despite their desire to create video content for more engaging storytelling, they struggled with limited video editing skills. 
With \SB{}, they found the process intuitive and began creating dynamic video stories to analyze player strengths and weaknesses in a more engaging format. Each video integrates multiple game clips,  breaking down insights with player and tactical visualizations to reinforce their main message, as illustrated in Fig.~\ref{fig:case_study}d. 

Overall, our collaborations with sports content creators of diverse needs and audiences demonstrated that \SB{} can effectively support their creative workflows and enhance video storytelling quality, whether by maintaining viewer engagement in long-format videos, clearly conveying team tactics and actions in short reels, or presenting compelling stories through detailed player analysis.

\section{Discussions}


\noindent\textbf{Bridge the gap between insights and expressions with AI-powered domain-focused video creation.}
As images and videos continue to dominate communication mediums, visualization and video technologies have become essential tools for enabling diverse domains and the public to express themselves effectively. Emerging generative AI tools, such as Sora~\cite{sora} and Pika~\cite{pika}, exemplify this trend by facilitating creative expression across various fields.

While general AI-driven video creation tools are increasingly popular, our work emphasizes the critical need for domain-specific video creation tools like \SB{} to address unique requirements within specific fields. There are two primary reasons for prioritizing domain-specific video creation over general generative technologies.
First, domain-specific videos, such as sports highlights, rely heavily on human insights. Audiences seek to learn from professionals through these videos, requiring tools that provide greater user control and enable experts to effectively translate their insights into engaging content. 
%
Second, the complexity of domain-specific data, such as the intricate motion and strategy analysis, demands advanced data visualization and seamless synchronization of visuals and audio, which general tools may not provide. 

\SB{} addresses these needs by integrating automation with customizable visualizations, tailored to the intricate and dynamic nature of sports content. It allows flexible user control through embedded interactions, 
reducing technical barriers and empowering users to effectively communicate their insights. Feedback from users further underscores the importance of balancing automation with user control to accommodate diverse goals and preferences to enhance accessibility across various user groups and use cases, such as tactical analysis, skill development, and profile building. 

Beyond sports, similar tools have the potential to transform fields like healthcare and education, incorporating precise visual aids and step-by-step breakdowns. 
%
Future research is required to investigate the balanced integration of AI and intuitive interface design, such as multi-modal interaction~\cite{wang2024lave}, to further advance domain-specific video creation and expression across diverse fields.



\vspace{1mm}
\noindent\textbf{Promote visualization in practice through real-world system deployment.}
Our work on SportsBuddy advances existing research in sports visualization and video authoring by emphasizing real-world system deployment and evaluation. Through this study, we have identified two significant benefits.

First, deploying SportsBuddy in authentic environments allowed us to validate and refine our design based on genuine use cases and users, uncovering insights that controlled laboratory settings cannot capture. For instance, we discovered that even within a similar user group of content creators, priorities varied significantly—some focused on showcasing player actions, while others emphasized strategic communication. This diversity led to iterative design improvements that balanced the distinct needs of each user group and support customization without complicating user interactions. 

Second, real-world deployment enables the assessment of long-term impacts and the discovery of unique use cases by diverse users. 
For example, some sports experts were hesitant to adopt SportsBuddy initially despite the perceived usefulness they shared. Upon further investigation, this was due to the context-switching costs. This feedback highlighted the necessity for a streamlined workflow tailored to the sports domain, leading to our design of batch processing and web import options. In addition, we observed many users preferred embedded annotation with \Text{} features over typical captions for sharing insights (see Fig.~\ref{fig:case_study}d), suggesting a new form of video storytelling inspired by \SB{}’s design. 
Feedback and insights from our diverse user base has highlighted the value of creating flexible and accessible visualization tools, which offers important external validity of the human-centered system.

This real-world deployment approach not only enhances visualization literacy and accessibility but also ensures that innovative designs translate into practical, widely usable tools, providing a validation for interactive visualization design. Therefore, we advocate for more visualization research to focus on real-world system deployments and to share design learnings, inspiring use cases that are both practical and impactful.

{
\subsection{Future Work}

While SportsBuddy has shown great potential in simplifying sports video storytelling, 
there are key areas for further improvement:

\vspace{1mm}
\noindent\textbf{Enhancing Player Tracking Under Occlusion and Motion Changes.}
The current tracking system faces challenges with occlusions and rapid motion in dynamic scenarios. Future work will refine tracking algorithms using larger domain-specific datasets and multi-view setups to improve accuracy in complex environments.


\vspace{1mm}
\noindent\textbf{Addressing Perspective and Camera Movement.}
Shifts in camera angles or perspectives cause misalignment issues due to reliance on fixed transformation matrices. Dynamic court mapping and machine learning for real-time adjustments, along with camera metadata integration, will ensure consistent and accurate visualizations.


\vspace{1mm}
\noindent\textbf{Supporting Longer Videos.}
Longer or higher-resolution videos can strain browser performance. To mitigate this, we will implement dynamic video loading from cloud storage and on-demand decoding, and adopt frame compression during previews to further optimize memory usage and rendering, ensuring smoother video processing.

\vspace{1mm}
\noindent\textbf{Extending to Other Sports.}
\SB{} currently focuses on basketball but can expand to sports like soccer and tennis. This requires adapting tracking algorithms and designing sport-specific visualizations to accommodate the unique dynamics and storytelling needs of each sport.

}


\section{Conclusion}
We designed and deployed \SB{}, an interactive video authoring tool that empowers effective sports storytelling. \SB{} features automatic player tracking and customizable highlight functionalities tailored for sports insight communication, allowing seamless integration of visual highlights with narratives through intuitive embedded interactions.  
During a three-month deployment, \SB{} attracted over 150 users with diverse roles and objectives in sharing personalized insights through video highlights. User feedback and case studies confirmed the tool’s usefulness and impact, resulting in the creation of engaging and creative video content. 
Our findings demonstrate the significant potential of intelligent, domain-focused video creation tools in sports and advocate for the broader adoption of similar visualization applications in real-world settings. 
\scriptsize
\acknowledgments{We wish to thank Nick Dow and Mike Sotsky for their invaluable support. Our gratitude also extends to HungKu, Hinbasket, JNC Sport, DP Women Basket, Boki Buckets, Dub Analysis, Thousand Basketball and all participants for their contributions. Additionally, we thank the reviewers for their valuable comments. }


\bibliographystyle{abbrv-doi}

\bibliography{template}
\end{document}